\documentclass[pra, 10pt, twocolumn, letter, floatfix]{revtex4-1}

\usepackage{amsmath}
\usepackage{amsfonts}
\usepackage{amssymb}
\usepackage{graphicx}
\usepackage{bm}
\usepackage{subcaption}

\begin{document}

\title[Blocked Linear Method] {A Blocked Linear Method for Optimizing Large Parameter Sets \\ in Variational Monte Carlo}

\author{Luning Zhao$^{1}$}
\author{Eric Neuscamman$^{1,2,}$}
\email{eneuscamman@berkeley.edu}
\affiliation{$^1$Department of Chemistry, University of California, Berkeley, California 94720, USA\\
             $^2$Chemical Sciences Division, Lawrence Berkeley National Laboratory, Berkeley, California 94720, USA}

\date{\today}

\begin{abstract}
We present a modification to variational Monte Carlo's linear method optimization scheme that addresses a critical memory bottleneck while
maintaining compatibility with both the traditional ground state variational principle and our recently-introduced variational
principle for excited states.
For wave function ansatzes with tens of thousands of variables, our modification reduces the required memory per parallel process
from tens of gigabytes to hundreds of megabytes, making the methodology a much better fit for modern supercomputer
architectures in which data communication and per-process memory consumption are primary concerns.
We verify the efficacy of the new optimization scheme in small molecule tests involving both the Hilbert space Jastrow antisymmetric
geminal power ansatz and real space multi-Slater Jastrow expansions.
Satisfied with its performance, we have added the optimizer to the QMCPACK software package, with which we demonstrate on a hydrogen ring
a prototype approach for making systematically convergent, non-perturbative predictions of Mott-insulators' optical band gaps.
\end{abstract}

\maketitle

\section{Introduction}
\label{sec:introduction}

In the ansatz-based approach to electronic structure theory, the capabilities of the method used to optimize the ansatz for
a particular system are every bit as important as the flexibility of the ansatz itself.
For example, both the coupled cluster \cite{Bartlett:2007:cc_rev} and matrix product state \cite{Schollwock:2011gl}
ansatzes would be much less useful if we lacked the projected Schr\"{o}dinger equation and density matrix renormalization group
methods that allow us to optimize them efficiently.
To address unsolved problems in electronic structure --- such as
catalytic cycles in which many bonds are simultaneously rearranged \cite{Yanai:2013:oec},
double excitations in large $\pi$-conjugated molecules\cite{Parac:2003:failure_tddft}, and 
high-temperature superconductivity \cite{Subedi:2008:supercond,zaanen:2015:high_tc} ---
it is therefore essential that improvements to optimization methods be made alongside innovations in ansatz design.

In few areas is the need for improved optimization methods more pressing than in quantum Monte Carlo (QMC).
Until very recently, optimization methods in this area were limited to a few thousand variational parameters
when using a fully ab initio Hamiltonian, a constraint that holds back progress in a wide variety of areas.
In fixed-node projector Monte Carlo methods such as diffusion Monte Carlo (DMC)
\cite{Foulkes:2001:qmc_review,Metropolis:2012:dmc,Umrigar:2015:qmc},
the inability to systematically converge the trial function's nodal surface due to insufficiently flexible
ansatzes is responsible for both the fixed node error and the pseudopotential locality error, the latter of
which becomes acutely problematic in 3rd-row and heavier elements where the nonlocal part of the pseudopotential
cannot be ignored.
Even in variational Monte Carlo \cite{Foulkes:2001:qmc_review,Umrigar:2015:qmc} (VMC) itself, recent innovations
in ansatz design create a pressing need for expanding the number of variational parameters that can be treated.
Examples in this category include the variation after response approach to excited states \cite{Neuscamman:2016:var},
efficient methods for large multi-Slater Jastrow (MSJ) expansions \cite{miguel:2011:table,Scuseria:2012:msj},
variational analogues of coupled cluster theory \cite{Zhao:2016:padccd},
and wave function stenciling approaches \cite{Neuscamman:2016:subtractive,Eric:2016:ScO,Booth:2016:opt} that tightly
couple the optimization of correlation factors and molecular orbitals.
For all of these reasons, and indeed for the simple reason of enabling systematic improvability within
a given ansatz, improvements in VMC optimization capabilities are sorely needed.

The linear method \cite{Nightingale:2001:linear_method,UmrTouFilSorHen-PRL-07,TouUmr-JCP-07,TouUmr-JCP-08} (LM) developed by Umrigar and coworkers
is currently the most effective VMC optimizer for cases in which the number of variables is a few thousand or less.
By solving a projected Schr\"{o}dinger equation in the vector space spanned by the current wave function and its first parameter derivatives,
a space we will refer to as the self-plus-tangent space, the LM produces update steps that account for second order couplings
between variables and in practice often out-perform Newton-Raphson steps, a success due in no small part to the fact that
these updates satisfy a strong zero variance principle \cite{TouUmr-JCP-07,TouUmr-JCP-08}.
However, the traditional LM's need to explicitly construct the Hamiltonian and overlap matrices in the self-plus-tangent space becomes
cumbersome when the number of variational parameters exceeds a few thousand due to the large amounts of memory required to store these matrices.
This issue becomes especially fraught when trying to match the LM to modern supercomputing resources, as each parallel Markov chain
must make space for its own copies of these matrices (a tall order given typical per-core memory restrictions), which must then be
communicated and combined prior to diagonalization.
While one could use Krylov subspace methods to solve the eigenproblem without explicitly constructing the matrices,
as was done for the related stochastic reconfiguration method \cite{Chan:2012:large_opt},
our experience in practice has taught us that finding a preconditioning scheme capable of reducing the condition
numbers of the LM matrices to manageable levels is not trivial.
As far as we are aware, these various issues have prevented the LM from being used in regimes beyond about 16,000 variables,
which occurred in the context of a ground state MSJ expansion for the water molecule \cite{miguel:2011:table}.

Very recently, Booth and coworkers introduced an alternative VMC optimization method that takes advantage of
optimal descent theory and a stochastic gradient evaluation in order to produce robust energy minimizations
despite avoiding second-derivatives entirely \cite{Booth:2016:opt}.
Impressively, this method appears capable of handling more than 60,000 variational parameters for ansatzes
that support efficient inner products with the Slater determinant basis functions of Fock space.
However, as the method relies on having fast access to Hamiltonian matrix elements between basis functions,
it is not immediately obvious how to extend it to the delta-function basis of real space where such matrix
elements are ill-defined.
Nonetheless, promising new directions in VMC optimization are a welcome development.

In the present study, we seek to retain the advantages of the traditional LM --- which include Fock space and real space
compatibility, robust convergence in a small number of iterations, and access to excited states through our recently
introduced \cite{Eric:2016:dir_tar} excited state variational principle --- while reducing its memory footprint so as
to facilitate larger variable sets and better compatibility with modern parallel computers.
Our strategy will be to separate the variable space into blocks, within each of which we estimate a small number of
important update directions that can then be used to construct a relatively small LM eigenproblem in the
overall basis of important directions.
We will demonstrate that this approach drastically reduces memory requirements without significantly affecting
the accuracy of the optimization.
In addition to tests on small molecules using our in-house Hilbert space software, we will use the implementation that
we recently contributed to the open-source QMCPACK software package \cite{qmcpack_1,qmcpack_2} to demonstrate this method's
excited state capabilities in the context of a hydrogen ring's Mott-like metal-insulator transition.
By evaluating the optical gap for a series of increasingly flexible MSJ expansions, the largest of which contains over 25,000
variational parameters, this study points the way towards a systematically convergent and non-perturbative 
approach to predicting optical gaps in the Mott-insulating regimes of real materials.

\section{Theory}
\label{sec:theory}

\subsection{The Linear Method}
\label{sec::lm}

The traditional LM works by repeatedly solving the Schr\"{o}dinger equation in the self-plus-tangent subspace of
the full Hilbert space, defined by the span of the wave function and its first derivatives with respect to its
variational parameters.
As the derivatives are not necessarily orthogonal to each other, this approach leads to a generalized eigenvalue problem 
\begin{align}
\label{eqn:eigen}
\sum _{ y\in \left\{ 0,1,... \right\}  }^{  }{ \left< { { \Psi  }^{ x } }|{ H }|{ { \Psi  }^{ y } } \right> { c }_{ y } } =\lambda \sum _{ y\in \left\{ 0,1,... \right\}  }^{  }{ \left< { { \Psi  }^{ x } }|{ { \Psi  }^{ y } } \right> { c }_{ y } } 
\end{align}
where $\left| \Psi^x \right>$ and $\left| \Psi^y \right>$ are the derivatives of $\left| \Psi \right>$ with respect to the $x$th and $y$th
wave function parameters $\mu_x$ and $\mu_y$, respectively, and $\left |{ \Psi  }^{ 0 } \right> \equiv \left |\Psi  \right> $.
After solving this eigenvalue problem for $\vec{c}$, one updates the parameters by
\begin{align}
\label{eqn:param_update}
{ \mu  }_{ x }\rightarrow { \mu  }_{ x }+{ c }_{ x }/{ c }_{ 0 } \quad \forall \quad x \in \left\{1,2,...\right\}
\end{align}
after which the updated $\left|\Psi\right>$ will be a good approximation for the subspace eigenfunction $\sum_y c_y \left|\Psi^y\right>$
so long as the updates $c_x/c_0$ are sufficiently small in magnitude.
This requirement can be ensured by applying a diagonal shift to the Hamiltonian matrix \cite{TouUmr-JCP-08}, which plays the same
role as a trust radius would in a Newton-Raphson optimization.
The updated ansatz in hand, a new self-plus-tangent space may be constructed and the procedure repeated until convergence is reached.  

In practice, the Hamiltonian and overlap matrix elements are estimated via Monte Carlo sampling,
\begin{align}
\label{eqn:matrix_build}
\begin{split}
&\sum _{ \vec { n }\in \xi  }^{  }{ \sum _{ y\in \left\{ 0,1,... \right\}  }^{  }
{ \frac{|\left<\vec{n}|\Psi\right>|^2}{\mathcal{P}(\vec{n})}
\frac { \left< { { \Psi  }^{ x } }|{ \vec { n } } \right>  }{ \left< { \Psi  }|{ \vec { n } } \right>  } \frac { \left< { \vec { n } }|{ H }|{ { \Psi  }^{ y } } \right>  }{ \left< { \vec { n } }|{ \Psi  } \right>  } { c }_{ y } }  } \\
&=\lambda \sum _{ \vec { n }\in \xi  }^{  }{ \sum _{ y\in \left\{ 0,1,... \right\}  }^{  }
{ \frac{|\left<\vec{n}|\Psi\right>|^2}{\mathcal{P}(\vec{n})}
\frac { \left< { { \Psi  }^{ x } }|{ \vec { n } } \right>  }{ \left< { \Psi  }|{ \vec { n } } \right>  } \frac { \left< { \vec { n } }|{ { \Psi  }^{ y } } \right>  }{ \left< { \vec { n } }|{ \Psi  } \right>  } { c }_{ y } }  }
\end{split}
\end{align}
where $\xi$ is a set of samples drawn from the probability distribution $\mathcal{P}(\vec{n})$ (which is typically chosen as
$|\left<\vec{n}|\Psi\right>|^2$) using Markov chain Monte Carlo.
Note that although we have depicted the sampling as running over occupation-number-vector-labeled determinants in Fock space,
the LM is equally viable if instead the sampling is carried out in real space, where $\mathcal{P}(\vec{r})$ is typically
chosen to be $|\Psi(\vec{r})|^2$.
The LM will thus be efficient (i.e.\ polynomial cost) for ansatzes that support the efficient evaluation of the derivative
ratios $\left<\vec{n}|\Psi^x\right>/\left<\vec{n}|\Psi\right>$ and $\left<\vec{n}|H|\Psi^x\right>/\left<\vec{n}|\Psi\right>$,
examples of which include MSJ expansions \cite{UmrTouFilSorHen-PRL-07,miguel:2011:table,Scuseria:2012:msj},
the Jastrow antisymmetric geminal power \cite{Sorella:2003:jagp,Sorella:2009:jagp_molec,Neuscamman:2013:hilbert_jagp} (JAGP),
and amplitude determinant coupled cluster with pairwise doubles \cite{Zhao:2016:padccd}.

While the cost scaling may be polynomial with system size, the memory required to store the Hamiltonian and overlap matrices
in the self-plus-tangent space can be a serious impediment to practical computation.
For example, when using 8-byte floating point numbers and an ansatz with 30,000 variational parameters,
the traditional LM requires 14.4 gigabytes of memory per Markov chain.
Such storage requirements create problems with the typical parallelization scheme of running one Markov chain per
core, as modern supercomputers typically have closer to 2 gigabytes of memory available per core.

One approach to circumventing matrix storage difficulties would be to use a Krylov subspace method to solve
for $\vec{c}$ without constructing the matrices explicitly.
While this strategy has shown promise in the related stochastic reconfiguration method, where it succeeded in
working with an ansatz containing half a million variables \cite{Chan:2012:large_opt}, Krylov subspace methods
are only efficient if the condition numbers of the matrices involved (the ratio of the magnitudes of their
largest and smallest magnitude eigenvectors) can be brought close to unity through preconditioning.
Although we have made some ad-hoc investigations into this area, we have not found preconditioners that
can reliably reduce the condition numbers involved below about $10^{10}$.
While this does not preclude the existence of an effective preconditioning scheme, it does prompt us to
investigate approaches, like the one in the next section, that remain effective even in the face of
highly ill-conditioned matrices.

\subsection{The Blocked Linear Method}
\label{sec::blm}

Ultimately, the goal of the LM is to find the best update direction and step length within the tangent space of the wave function.
Imagine instead holding half the variables fixed and inspecting the tangent space for the other half.
The diagonalization of the linear method eigenproblem within this self-plus-half-tangent space will produce a set of update directions
that can be ordered by importance, as measured by their eigenvalues, which inform us as to how much a move along an eigen-direction
would decrease or increase the energy.
Noting that the optimal direction $\vec{c}_{\mathrm{opt}}$ in the full tangent space, whose dimension is the total number of variational
parameters $N_V$, can be written as a linear combination of $N_V/2$ orthogonal directions within one half-tangent space and $N_V/2$ orthogonal
directions from the other half-tangent space, it seems intuitive that a very bad update direction in one of the half-tangent spaces
is unlikely to be an important component of $\vec{c}_{\mathrm{opt}}$.
Taken further, this logic suggests that it may be possible to construct a close approximation to $\vec{c}_{\mathrm{opt}}$ using a linear
combination of only a few update directions from each half-tangent space.
In essence, the blocked linear method (BLM) is an attempt to systematically exploit this structure by (a) dividing the variable space
into a number of blocks, (b) making intelligent estimates for which directions within those blocks will be most important for
constructing $\vec{c}_{\mathrm{opt}}$, and (c) estimating $\vec{c}_{\mathrm{opt}}$ by solving a smaller, more memory-efficient
eigenproblem in the basis of these supposedly important block-wise directions.

Rather than the traditional LM's expansion of the wave function in its self-plus-tangent space, consider instead the ``one-block'' expansion
\begin{align}
\label{eqn:blm_expansion}
|\Phi_b\rangle = \alpha_b |\Psi\rangle
 + \sum_{i=1}^{M_b} \beta_{bi} |\Psi^{i,b}\rangle
 + \sum_{j=1}^{N_O} \sum_{\substack{k=1 \\ k \neq b}}^{N_B} \gamma_{bjk} |\Theta_{jk}\rangle.
\end{align}
In the first two terms, we have a linear expansion of the wave function with respect
to the variables belonging to the $b$th block,
with $\alpha_b$ and $\beta_{bi}$ the expansion coefficients,
$M_b$ the number of variables in the block,
and $|\Psi^{i,b}\rangle$ defined as the wave function derivative with respect to the $i$th variable of the $b$th block.
If we drop the third term for now (i.e.\ set $\gamma_{bjk}=0$), we have a wave function whose energy minimization
\begin{align}
\label{eqn:block_min}
\substack{\mathrm{min} \\ \alpha, \beta} \hspace{2mm}
\langle\Phi_b| \hat{H} |\Phi_b\rangle / \langle\Phi_b|\Phi_b\rangle
\end{align}
leads to a generalized eigenvalue problem in the same form as for the traditional LM, Eq.\ (\ref{eqn:eigen}),
the only difference being that we are now holding the variables outside the chosen block fixed.
(Note that while we will develop the discussion here in terms of energy minimization, the BLM is equally applicable to the target function
used in the direct, variational targeting of excited states \cite{Eric:2016:dir_tar} and has been implemented and tested for both cases).
Each eigenvector will have its own values for the $\alpha_b$ and $\beta_{bi}$ coefficients and will correspond to an eigenvalue
that gives an estimate for what the energy of our original wave function would be if we were to update this block's variables 
according to $\mu_{i,b}\rightarrow\mu_{i,b} + \beta_{bi} / \alpha_b$.
Thus, the eigenvalues of this block's eigenproblem inform us as to which directions in its variable space are expected to
be ``good'' update directions (those with the lowest eigenvalues) and which are expected to be ``bad'' directions (those with the highest eigenvalues).

Having performed this diagonalization within each of our blocks, we are now in a position to construct an approximation
to the wave function in its full self-plus-tangent space by retaining from each variable block only a small number of what are expected
to be the best update directions.
By organizing the best $N_K$ update directions from the $b$th block
into the rows of a matrix $\bm{B}^{(b)}$, this self-plus-tangent space approximation can be written as
\begin{align}
\label{eqn:approx_in_full_spt}
|\Upsilon(\alpha,\bm{A})\rangle = \alpha |\Psi\rangle
 + \sum_{b=1}^{N_B} \sum_{j=1}^{N_K} A_{bj} \sum_{i=1}^{M_b} B^{(b)}_{ji} |\Psi^{i,b}\rangle.
\end{align}
As the elements of the $\bm{B}$ matrices are now held fixed, this expansion is not as flexible as that of the traditional LM,
but we hope the fact that it is built out of a linear combination of the best update directions from each block
will give it the correct flexibility to closely approximate the optimal update direction in the full tangent space.
This direction is now estimated via
\begin{align}
\label{eqn:full_min}
\substack{\mathrm{min} \\ \alpha, \bm{A}} \hspace{2mm}
\langle\Upsilon| \hat{H} |\Upsilon\rangle / \langle\Upsilon|\Upsilon\rangle
\end{align}
which again produces a generalized eigenvalue problem, this time of dimension $1+N_BN_K$, whose lowest
energy eigenvector corresponds to the overall BLM update,
\begin{align}
\label{eqn:final_blm_update}
\mu_{i,b} \rightarrow \mu_{i,b} + \frac{[ \bm{A} \bm{B}^{(b)} ]_{bi}}{\alpha}.
\end{align}
Crucially, the Hamiltonian and overlap matrix elements involved in the eigenvalue problems that stem from
Eqs.\ (\ref{eqn:block_min}) and (\ref{eqn:full_min}) can be estimated using the same information
as in the traditional LM, namely the derivative ratios 
$\left<\vec{n}|\Psi^x\right>/\left<\vec{n}|\Psi\right>$ and $\left<\vec{n}|H|\Psi^x\right>/\left<\vec{n}|\Psi\right>$,
at each sampled configuration $\vec{n}$ (or position $\vec{r}$ in real space).
While the most efficient way to construct these matrices now that the $\bm{B}^{(b)}$ coefficients
are known appears to be to re-run the same sample that was used to construct the block-specific
matrices, we feel that this second sampling is a price worth paying in order to remove the
traditional LM's memory bottleneck.

So far, we have ignored the fact that inter-block variable couplings will affect
which directions in a block are optimal for use in constructing an overall update direction.
Accounting for such couplings is the purpose of the third term in Eq.\ (\ref{eqn:blm_expansion}),
in which
\begin{align}
\label{eqn:previous_dir}
|\Theta_{jk}\rangle = \sum_{l=1}^{M_k} D_{jkl} |\Psi^{l,k}\rangle
\end{align}
is a linear combination of wave function derivatives from the $k$th block that is presumed to correspond
to a good update direction for that block.
By including a small number $N_O$ of these directions from each other block in the wave function expansion
$|\Phi_b\rangle$ for the current block, we hope to provide the minimization
\begin{align}
\label{eqn:full_block_min}
\substack{\mathrm{min} \\ \alpha, \beta, \gamma} \hspace{2mm}
\langle\Phi_b| \hat{H} |\Phi_b\rangle / \langle\Phi_b|\Phi_b\rangle,
\end{align}
which replaces that of Eq.\ (\ref{eqn:block_min}) in the overall method outlined above,
with the coupling information necessary so that the directions it contributes to $\bm{B}^{(b)}$
are optimal with respect to both intra-block and inter-block variable couplings.
While there are many possible choices for the linear combinations $|\Theta_{jk}\rangle$, we thought it
natural to derive them from previous iterations' BLM updates, following the idea that using
previous update directions to inform the current direction is a common theme in numerical minimization,
occurring for example in both the BFGS \cite{Nocedal:1980:lbfgs} and accelerated descent
\cite{Nesterov:1983:accel_descent} methods.
Specifically, for the $n$th iteration of the BLM, we take $|\Theta_{jk}\rangle$ as the $k$th block's
component of the $(n-j)$th iteration's overall update, with $j\in\{1, 2, \ldots, N_O\}$.
As our results will demonstrate, even relatively short history lengths $N_O$ can be
beneficial in accounting for inter-block variable couplings and thereby recovering the
performance of the traditional LM.

\begin{figure}[t]
\includegraphics[width=0.90\linewidth,angle=0]{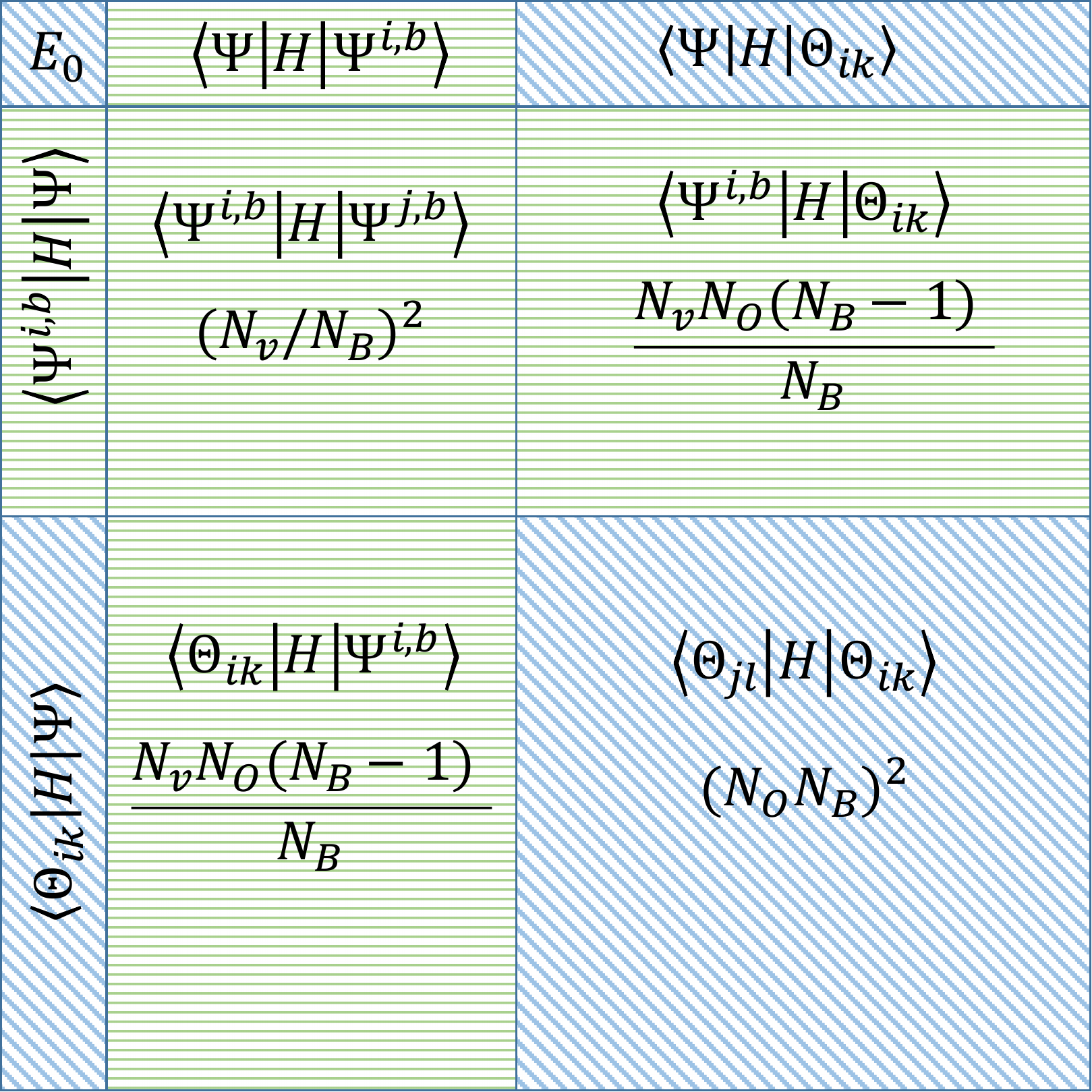}
\caption{Structure of the BLM Hamiltonian matrix for the $b$th block, with each section of the
         matrix displaying its type of matrix element.
         Green-shaded sections contain elements that are unique to each block; for the larger among
         these, we print the number of elements that must be stored per block.
         Blue-shaded sections contain elements shared by all blocks; for the larger among
         these, we print the total storage requirement across all blocks.
         Total memory consumption can then be evaluated as
         blue + $N_B\times$green.
        }
\label{fig:mat_struct}
\end{figure}

To understand the reduced memory footprint of the BLM, it is helpful to consult a visual guide to
the structure of the Hamiltonian and overlap matrices resulting from Eq.\ (\ref{eqn:full_block_min}).
Figure \ref{fig:mat_struct} shows this structure for the Hamiltonian; the overlap matrix
has an analogous structure.
Noting that the different blocks' eigenproblems can be solved independently, we can see that only one
block's matrices need to be fully constructed at a time, which greatly reduces memory requirements
by allowing us to store one copy, rather than $N_B$ copies, of the blue elements in Figure \ref{fig:mat_struct}.
For the green elements, however, we must store $N_B$ copies simultaneously, so that each sampled
configuration $\vec{n}$ or $\vec{r}$ can efficiently add its unique contribution to each of them.
Nonetheless, storage requirements are much lower than in the traditional LM, whose Hamiltonian
matrix contains $(1+N_V)^2$ elements.
Although the precise formula for the BLM's Hamiltonian storage requirement is more longwinded,
the terms that dominate,
$N_V^2/N_B$ and $2 N_V N_O (N_B-1)$, are much smaller than the dominant $N_V^2$ term in the LM.
Thus, if no previous updates are being used (i.e.\ $N_O=0$), the BLM reduces memory requirement by
a factor of $N_B$, and although the use of $N_O>0$ increases
the BLM's memory requirement somewhat, the savings remain substantial.
For example, when using 8-byte floats and 30,000 variational parameters,
the traditional LM requires 14.4 gigabytes of memory per process,
while the BLM with $N_B=100$ and $N_O=5$ requires only 0.5 gigabytes per process.

\section{Results}

\subsection{Computational Details}

JAGP results for N$_2$ and H$_2$O were obtained using Hilbert-space sampling via our own VMC software, which extracts
one- and two-electron integrals from PySCF \cite{Pyscf_brief}.
MSJ results for C$_2$ and the hydrogen ring were obtained using real-space sampling via QMCPACK \cite{qmcpack_1,qmcpack_2},
with configuration state functions (CSFs) taken from GAMESS \cite{gamess1}.
For JAGP, we work exclusively in the symmetrically orthogonalized ``$S^{-1/2}$'' one particle basis. 
The VMC sample size is universally chosen as 2.4$\times$10$^5$, which produces statistical uncertainties
whose standard deviations are less than 0.7 kcal/mol (0.03 eV) in all cases.

\subsection{N$_2$ and H$_2$O with JAGP}

\begin{table}
\caption{Comparison of the LM ($N_B=1$) and BLM for the ground state of N$_2$ using the JAGP ansatz with Hilbert-space sampling in the 6-31G basis.}
\label{tab:n2}
\begin{tabular}{ c c c r@{.}l r@{.}l c }
\hline\hline
\hspace{0mm} $N_B$ \hspace{0mm} & \hspace{0mm} $N_O$ \hspace{0mm} & \hspace{0mm} $N_K$ \hspace{0mm} & 
\multicolumn{2}{ c }{ \hspace{0mm} Energy (a.u.) \hspace{0mm} } &
\multicolumn{2}{ c }{ \hspace{0mm} Error (eV) \hspace{0mm} } &
\hspace{0mm} N$_{\mathrm{iter}}$ \hspace{0mm} \\
\hline
 \hspace{0mm} 1 \hspace{0mm} & \hspace{0mm} N/A \hspace{0mm} & \hspace{0mm} N/A \hspace{0mm} & \hspace{0mm} -109&089 \hspace{0mm} & \hspace{0mm} 0&00 \hspace{0mm} & \hspace{0mm} 18 \hspace{0mm} \\
 \hspace{0mm} 4 \hspace{0mm} & \hspace{0mm} 1 \hspace{0mm} & \hspace{0mm} 1 \hspace{0mm} & \hspace{0mm} -109&088 \hspace{0mm} & \hspace{0mm} 0&04 \hspace{0mm} & \hspace{0mm} 19 \hspace{0mm} \\
 \hspace{0mm} 4 \hspace{0mm} & \hspace{0mm} 5 \hspace{0mm} & \hspace{0mm} 1 \hspace{0mm} & \hspace{0mm} -109&088 \hspace{0mm} & \hspace{0mm} 0&04 \hspace{0mm} & \hspace{0mm} 21 \hspace{0mm} \\
 \hspace{0mm} 8 \hspace{0mm} & \hspace{0mm} 1 \hspace{0mm} & \hspace{0mm} 1 \hspace{0mm} & \hspace{0mm} -109&087 \hspace{0mm} & \hspace{0mm} 0&08 \hspace{0mm} & \hspace{0mm} 29 \hspace{0mm} \\
 \hspace{0mm} 8 \hspace{0mm} & \hspace{0mm} 5 \hspace{0mm} & \hspace{0mm} 1 \hspace{0mm} & \hspace{0mm} -109&087 \hspace{0mm} & \hspace{0mm} 0&08 \hspace{0mm} & \hspace{0mm} 19 \hspace{0mm} \\
 \hspace{0mm} 16 \hspace{0mm} & \hspace{0mm} 1 \hspace{0mm} & \hspace{0mm} 1 \hspace{0mm} & \hspace{0mm} -109&086 \hspace{0mm} & \hspace{0mm} 0&09 \hspace{0mm} & \hspace{0mm} 38 \hspace{0mm} \\
 \hspace{0mm} 16 \hspace{0mm} & \hspace{0mm} 5 \hspace{0mm} & \hspace{0mm} 1 \hspace{0mm} & \hspace{0mm} -109&086 \hspace{0mm} & \hspace{0mm} 0&09 \hspace{0mm} & \hspace{0mm} 24 \hspace{0mm} \\
\hline\hline
\end{tabular}
\end{table}

We begin our numerical tests with the ground states of two small molecules, N$_2$ and H$_2$O,
choosing the JAGP for our ansatz and performing VMC sampling in the second-quantized Hilbert-space of the 6-31G \cite{6-31g} orbital basis.
These choices give us 408 and 273 nonlinear parameters to optimize in N$_2$ and H$_2$O, respectively,
which are few enough so as to make direct comparisons to the traditional LM straightforward.
Tables \ref{tab:n2} and \ref{tab:h2o} show the results for various combinations of the number of blocks $N_B$, previous update vectors $N_O$,
and retained block directions $N_K$.
The reported optimization error is defined as the difference in energy 
between the minimums found by the BLM and the traditional LM, the latter of which is denoted by $N_B=1$ in the tables.

The first observation to be made is that although small, errors with respect to the traditional LM are not zero.
The likely explanation for this fact is that the BLM update direction, like that of the traditional LM, is a nonlinear
function of the random variables drawn by our Markov chains.
Unlike linear functions of random variables that have statistical uncertainty but no systematic bias, nonlinear functions produce
a systematic bias, albeit one that can in principle be mitigated by increasing the sample size.
We suspect that our 2-step process of first diagonalizing $N_B$ block-wise eigenproblems before constructing and diagonalizing
one overall eigenproblem, which we note uses the same VMC sample for both steps, is essentially more nonlinear than the
traditional method's 1-step process.
In other words, both the BLM and LM should be expected to converge to a point in variable space that is slightly off-center from
the true minimum due to systematic bias, but we expect the BLM to be more off-center due to its additional nonlinearities.
Indeed, we have verified that the two methods converge to the same minimum in the limit of infinite sampling,
and as can be seen in the results, differences for finite sample lengths are modest and decrease as we retain more directions
$N_K$ from each block.

\begin{table}[t]
\caption{Comparison of the LM ($N_B=1$) and BLM for the ground state of H$_2$O using the JAGP ansatz with Hilbert-space sampling in the 6-31G basis.}
\label{tab:h2o}
\begin{tabular}{ c c c r@{.}l r@{.}l c }
\hline\hline
\hspace{0mm} $N_B$ \hspace{0mm} & \hspace{0mm} $N_O$ \hspace{0mm} & \hspace{0mm} $N_K$ \hspace{0mm} & 
\multicolumn{2}{ c }{ \hspace{0mm} Energy (a.u.) \hspace{0mm} } &
\multicolumn{2}{ c }{ \hspace{0mm} Error (eV) \hspace{0mm} } &
\hspace{0mm} N$_{\mathrm{iter}}$ \hspace{0mm} \\
\hline
 \hspace{0mm} 1 \hspace{0mm} & \hspace{0mm} N/A \hspace{0mm} & \hspace{0mm} N/A \hspace{0mm} & \hspace{0mm} -76&109 \hspace{0mm} & \hspace{0mm} 0&00 \hspace{0mm} & \hspace{0mm} 8 \hspace{0mm} \\
 \hspace{0mm} 2 \hspace{0mm} & \hspace{0mm} 5 \hspace{0mm} & \hspace{0mm} 1 \hspace{0mm} & \hspace{0mm} -76&108 \hspace{0mm} & \hspace{0mm} 0&04 \hspace{0mm} & \hspace{0mm} 10 \hspace{0mm} \\
 \hspace{0mm} 4 \hspace{0mm} & \hspace{0mm} 1 \hspace{0mm} & \hspace{0mm} 1 \hspace{0mm} & \hspace{0mm} -76&106 \hspace{0mm} & \hspace{0mm} 0&08 \hspace{0mm} & \hspace{0mm} 11 \hspace{0mm} \\
 \hspace{0mm} 4 \hspace{0mm} & \hspace{0mm} 3 \hspace{0mm} & \hspace{0mm} 1 \hspace{0mm} & \hspace{0mm} -76&106 \hspace{0mm} & \hspace{0mm} 0&09 \hspace{0mm} & \hspace{0mm} 11 \hspace{0mm} \\
 \hspace{0mm} 4 \hspace{0mm} & \hspace{0mm} 5 \hspace{0mm} & \hspace{0mm} 1 \hspace{0mm} & \hspace{0mm} -76&106 \hspace{0mm} & \hspace{0mm} 0&08 \hspace{0mm} & \hspace{0mm} 11 \hspace{0mm}\\
 \hspace{0mm} 8 \hspace{0mm} & \hspace{0mm} 1 \hspace{0mm} & \hspace{0mm} 1 \hspace{0mm} & \hspace{0mm} -76&103 \hspace{0mm} & \hspace{0mm} 0&16 \hspace{0mm} & \hspace{0mm} 9 \hspace{0mm}\\
 \hspace{0mm} 8 \hspace{0mm} & \hspace{0mm} 1 \hspace{0mm} & \hspace{0mm} 2 \hspace{0mm} & \hspace{0mm} -76&104 \hspace{0mm} & \hspace{0mm} 0&13 \hspace{0mm} & \hspace{0mm} 10 \hspace{0mm}\\
 \hspace{0mm} 8 \hspace{0mm} & \hspace{0mm} 1 \hspace{0mm} & \hspace{0mm} 4 \hspace{0mm} & \hspace{0mm} -76&107 \hspace{0mm} & \hspace{0mm} 0&06 \hspace{0mm} & \hspace{0mm} 13 \hspace{0mm}\\
 \hspace{0mm} 8 \hspace{0mm} & \hspace{0mm} 3 \hspace{0mm} & \hspace{0mm} 1 \hspace{0mm} & \hspace{0mm} -76&104 \hspace{0mm} & \hspace{0mm} 0&13 \hspace{0mm} & \hspace{0mm} 12 \hspace{0mm}\\
 \hspace{0mm} 8 \hspace{0mm} & \hspace{0mm} 3 \hspace{0mm} & \hspace{0mm} 2 \hspace{0mm} & \hspace{0mm} -76&106 \hspace{0mm} & \hspace{0mm} 0&08 \hspace{0mm} & \hspace{0mm} 12 \hspace{0mm}\\
 \hspace{0mm} 8 \hspace{0mm} & \hspace{0mm} 3 \hspace{0mm} & \hspace{0mm} 4 \hspace{0mm} & \hspace{0mm} -76&106 \hspace{0mm} & \hspace{0mm} 0&08 \hspace{0mm} & \hspace{0mm} 11 \hspace{0mm}\\
 \hspace{0mm} 8 \hspace{0mm} & \hspace{0mm} 5 \hspace{0mm} & \hspace{0mm} 1 \hspace{0mm} & \hspace{0mm} -76&107 \hspace{0mm} & \hspace{0mm} 0&07 \hspace{0mm} & \hspace{0mm} 12 \hspace{0mm}\\
 \hspace{0mm} 8 \hspace{0mm} & \hspace{0mm} 5 \hspace{0mm} & \hspace{0mm} 2 \hspace{0mm} & \hspace{0mm} -76&106 \hspace{0mm} & \hspace{0mm} 0&08 \hspace{0mm} & \hspace{0mm} 13 \hspace{0mm}\\
 \hspace{0mm} 8 \hspace{0mm} & \hspace{0mm} 5 \hspace{0mm} & \hspace{0mm} 4 \hspace{0mm} & \hspace{0mm} -76&108 \hspace{0mm} & \hspace{0mm} 0&04 \hspace{0mm} & \hspace{0mm} 12 \hspace{0mm}\\
\hline\hline
\end{tabular}
\end{table}

\begin{table}[b]
\caption{Comparison of the LM ($N_B=1$) and BLM for the ground state of C$_2$ using a MSJ expansion with real-space sampling.}
\label{tab:c2}
\begin{tabular}{ c c c r@{.}l r@{.}l c }
\hline\hline
\hspace{0mm} $N_B$ \hspace{0mm} & \hspace{0mm} $N_O$ \hspace{0mm} & \hspace{0mm} $N_K$ \hspace{0mm} & 
\multicolumn{2}{ c }{ \hspace{0mm} Energy (Hartree) \hspace{0mm} } &
\multicolumn{2}{ c }{ \hspace{0mm} Error (eV) \hspace{0mm} } &
\hspace{0mm} N$_{\mathrm{iter}}$ \hspace{0mm} \\
\hline
 \hspace{0mm} 1 \hspace{0mm} & \hspace{0mm} N/A \hspace{0mm} & \hspace{0mm} N/A \hspace{0mm} & \hspace{0mm} -75&834 \hspace{0mm} & \hspace{0mm} 0&00 \hspace{0mm} & \hspace{0mm} 8 \hspace{0mm}\\
 \hspace{0mm} 4 \hspace{0mm} & \hspace{0mm} 1 \hspace{0mm} & \hspace{0mm} 1 \hspace{0mm} & \hspace{0mm} -75&834 \hspace{0mm} & \hspace{0mm} 0&01 \hspace{0mm} & \hspace{0mm} 8 \hspace{0mm}\\
 \hspace{0mm} 8 \hspace{0mm} & \hspace{0mm} 1 \hspace{0mm} & \hspace{0mm} 1 \hspace{0mm} & \hspace{0mm} -75&833 \hspace{0mm} & \hspace{0mm} 0&03 \hspace{0mm} & \hspace{0mm} 10 \hspace{0mm}\\
 \hspace{0mm} 16 \hspace{0mm} & \hspace{0mm} 1 \hspace{0mm} & \hspace{0mm} 1 \hspace{0mm} & \hspace{0mm} -75&833 \hspace{0mm} & \hspace{0mm} 0&01 \hspace{0mm} & \hspace{0mm} 11 \hspace{0mm}\\
 \hspace{0mm} 50 \hspace{0mm} & \hspace{0mm} 1 \hspace{0mm} & \hspace{0mm} 1 \hspace{0mm} & \hspace{0mm} -75&832 \hspace{0mm} & \hspace{0mm} 0&04 \hspace{0mm} & \hspace{0mm} 10 \hspace{0mm}\\
 \hspace{0mm} 100 \hspace{0mm} & \hspace{0mm} 1 \hspace{0mm} & \hspace{0mm} 1 \hspace{0mm} & \hspace{0mm} -75&827 \hspace{0mm} & \hspace{0mm} 0&18 \hspace{0mm} & \hspace{0mm} 12 \hspace{0mm}\\
 \hspace{0mm} 100 \hspace{0mm} & \hspace{0mm} 5 \hspace{0mm} & \hspace{0mm} 1 \hspace{0mm} & \hspace{0mm} -75&831 \hspace{0mm} & \hspace{0mm} 0&08 \hspace{0mm} & \hspace{0mm} 11 \hspace{0mm}\\
 \hspace{0mm} 100 \hspace{0mm} & \hspace{0mm} 5 \hspace{0mm} & \hspace{0mm} 5 \hspace{0mm} & \hspace{0mm} -75&832 \hspace{0mm} & \hspace{0mm} 0&04 \hspace{0mm} & \hspace{0mm} 10 \hspace{0mm}\\
\hline\hline
\end{tabular}
\end{table}

\begin{figure*}
\begin{subfigure}{0.49\linewidth}
\centering
\includegraphics[width=0.80\linewidth,angle=270]{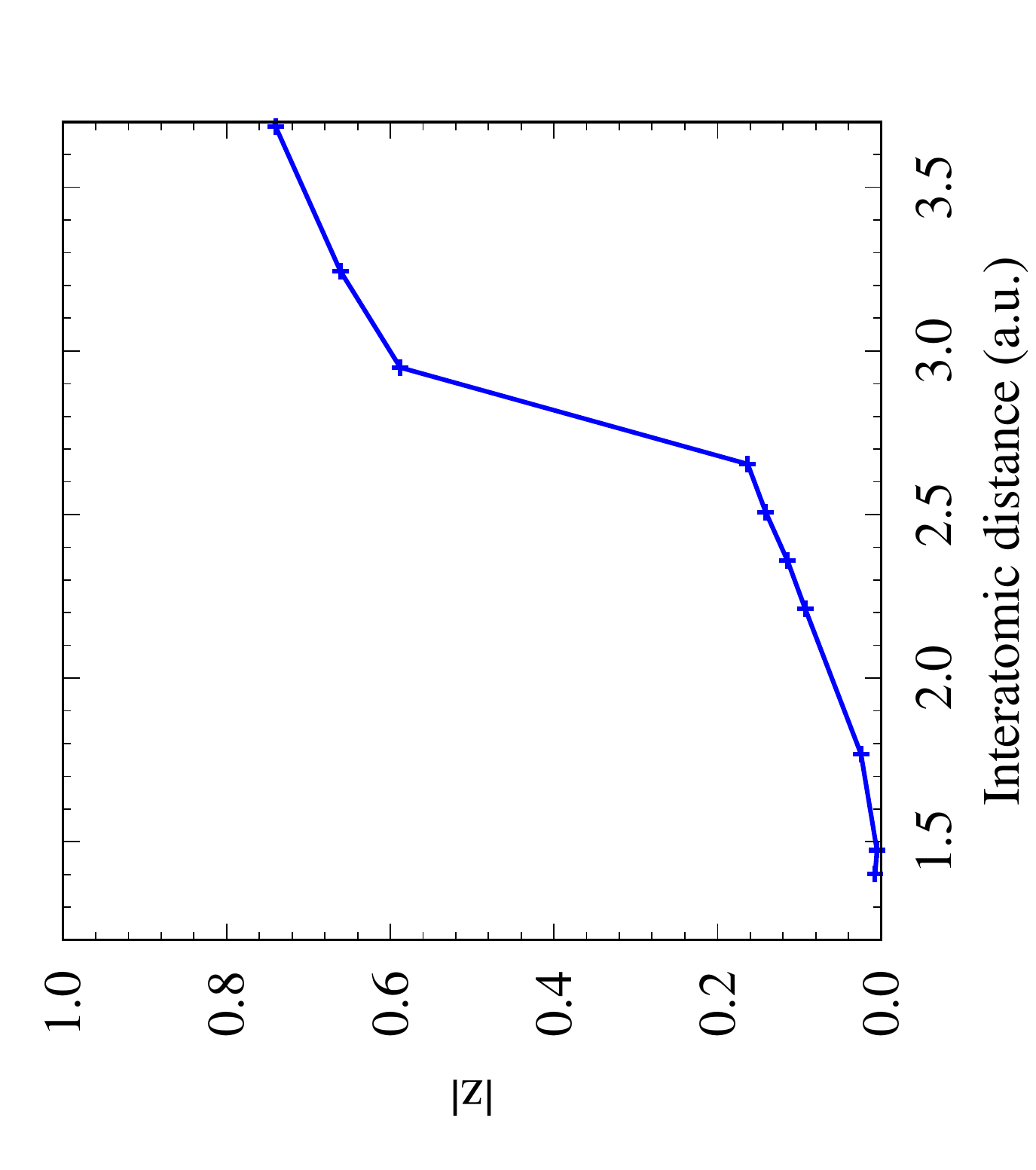}
\end{subfigure}\hfill
\begin{subfigure}{0.49\linewidth}
\centering
\includegraphics[width=0.80\linewidth,angle=270]{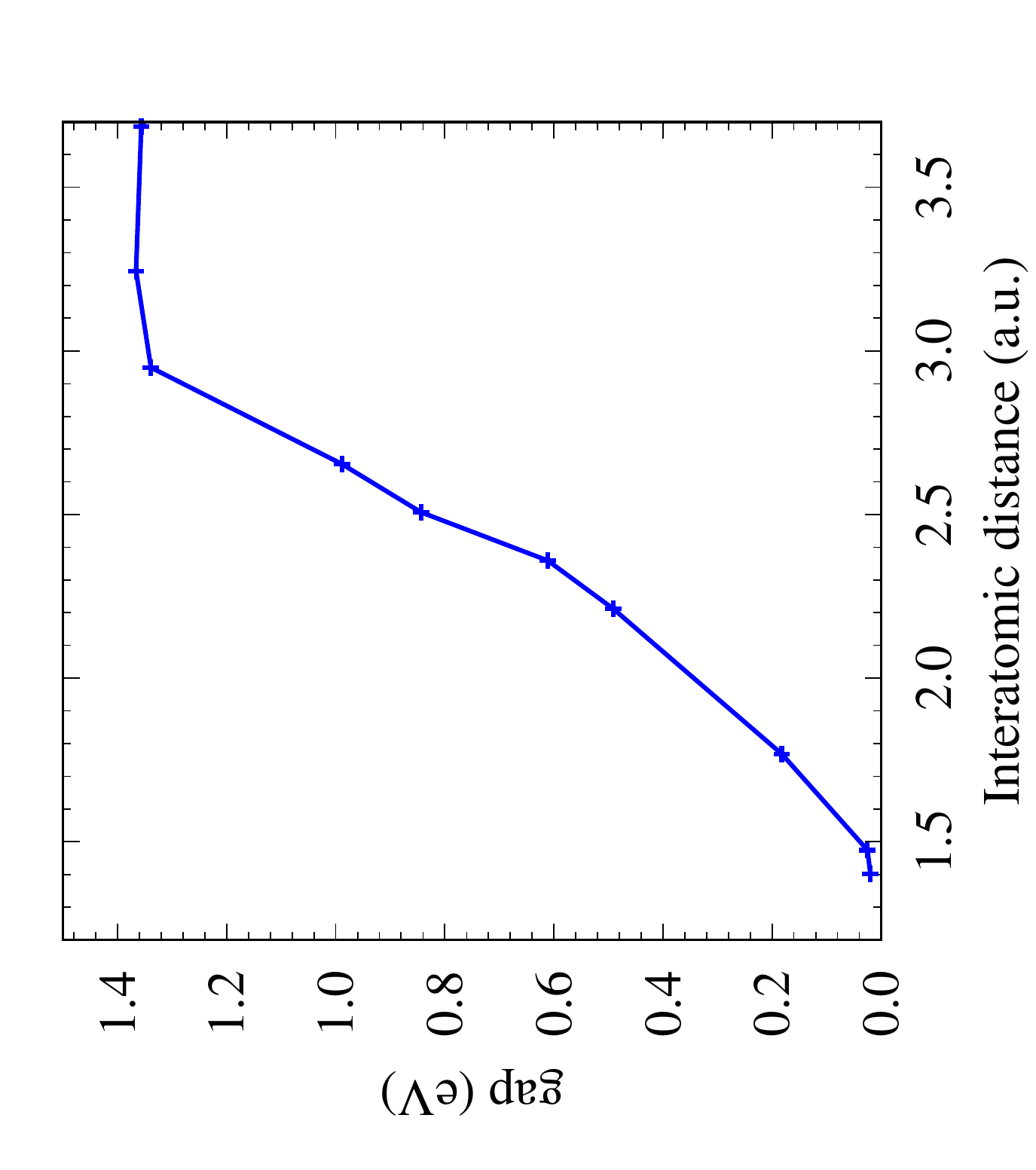}
\end{subfigure}\hfill
\caption{The complex polarization $|z|$ and optical gap of the H$_{16}$ ring as a function of the interatomic distance,
         evaluated using a MSJ ansatz containing all CISDTQ configurations with coefficients above 0.01.}
\label{fig:z_and_gap}
\end{figure*}

The second observation is that the error behaves as expected for different values of $N_B$, $N_K$, and $N_O$.
Increasing the number of blocks $N_B$, which makes it harder to account for second-order couplings between variables when
choosing update directions, increases the deviation from the traditional LM.
Also as expected, increasing $N_K$ and $N_O$ tends to decrease the deviation.
As hypothesized in the motivation for the BLM, only modest values of $N_K$ and $N_O$ are required to
produce close approximations to the optimal update direction, and so mitigating
deviations from the traditional LM is not difficult.
Finally, we note that although the BLM typically requires more iterations to converge,
the convergence speed remains similar to the traditional LM, especially when taking advantage of both
multiple directions $N_K$ per block and some number $N_O$ of previous update directions.

\subsection{C$_2$ with MSJ}

We next switch from sampling in Fock space to sampling in real space, with Table \ref{tab:c2} giving results for the ground state of C$_2$ as modeled by a MSJ
ansatz containing 1,100 CSFs and 30 spline-based Jastrow variables.
To construct our CSF expansion, we began with a GAMESS optimization of an (8,8) complete active space self-consistent field (CASSCF) ansatz
in the cc-pVTZ basis \cite{cc-pvdz}.
The 1,100 largest-coefficient CSFs were then selected from a single-reference configuration interaction calculation including up to quadruples (CISDTQ)
performed in the optimized CASSCF orbital basis.
As before, we see that increasing the number of blocks eventually results in a significant deviation from the traditional LM energy,
which is then reduced by increasing the number of old updates used and the number of directions retained from each block.
Again, while larger, the number of iterations required to converge the BLM was similar to that for the traditional LM.

\begin{figure*}
\begin{subfigure}{0.49\linewidth}
\centering
\includegraphics[width=0.87\linewidth,angle=270]{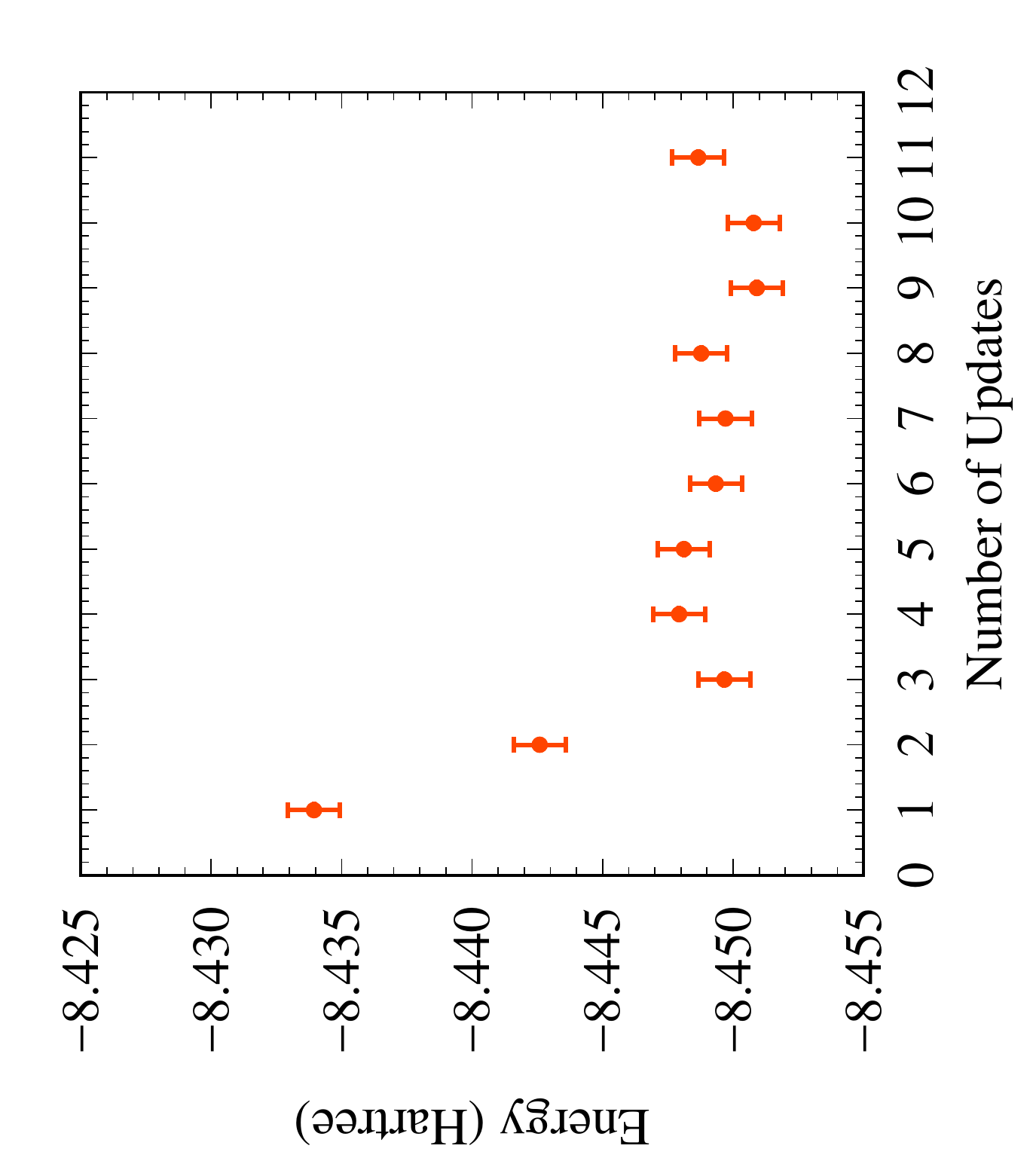}
\end{subfigure}\hfill
\begin{subfigure}{0.49\linewidth}
\centering
\includegraphics[width=0.87\linewidth,angle=270]{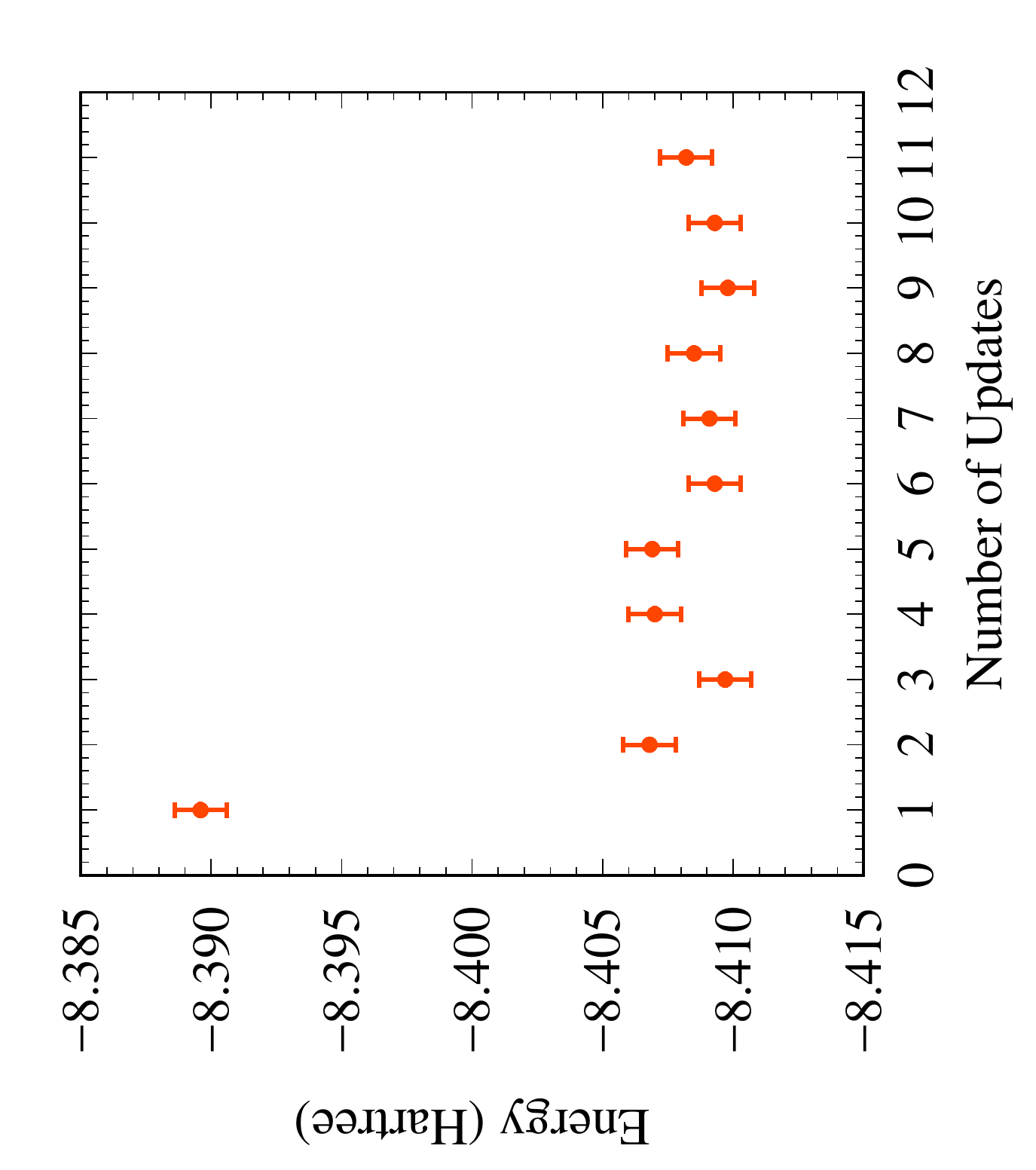}
\end{subfigure}\hfill
\caption{BLM convergence for the hydrogen ring's MSJ energy in the ground state (left, 21,401 parameters)
         and first excited state (right, 25,297 parameters).
        }
\label{fig:energy_convergence}
\end{figure*}

\subsection{The H$_{16}$ Hydrogen Ring}

Having tested our method in settings where it can be easily checked against the traditional LM, we now turn our attention to
the metal-insulator transition in a 16-atom hydrogen ring, where we will use the BLM in conjunction with our excited state targeting
method \cite{Eric:2016:dir_tar} to systematically converge the post-transition optical gap via a series
of increasingly large MSJ expansions.
Closely related hydrogen chains have been the subject of much attention \cite{Chan:2006:h_chain,Gus:2009:h_chain,sorella:2011:h_chain}
due to the Mott-like behavior of the metal-insulator transition that occurs as one enlarges the interatomic distance $a$.
As $a$ surpasses a certain critical distance $a_c$, a large number of natural orbitals become degenerate as the electrons transition
out of the weakly correlated metallic state and into the strongly correlated and more localized Mott-insulator state.

Using JAGP approximations for the ground state of the 1D chain, Sorrela and coworkers \cite{sorella:2011:h_chain} located $a_c$ by evaluating
the complex polarization function \cite{Marzari:2005:z}
\begin{align}
\label{eqn:comp_pl}
z = \langle\Psi| \hspace{1mm} \mathrm{exp} \left( \frac{2 \pi i}{L} \sum_k r^{\parallel}_k \right) |\Psi\rangle,
\end{align}
where $r^{\parallel}_k$ is the component of $\vec{r}_k$ parallel to the chain axis. 
The modulus of $z$ can be thought of as a measurement of insulating behavior:
$|z|\rightarrow1$ as electrons localize about the nuclei, as occurs in the insulating phase, 
while $|z|\rightarrow0$ as the electrons become fully delocalized, as occurs in the metallic phase.
As we are studying a hydrogen ring instead of a periodic chain, we find it appropriate to
instead define the complex polarization function as
\begin{align}
\label{eqn:comp_pl_ring}
z = \langle\Psi| \hspace{1mm} \mathrm{exp} \left( i \sum_k \theta_k \right) |\Psi\rangle,
\end{align}
where $\theta_k$ is the angle around the ring for the $k$th electron's position.
As for the chain, fully localized versus delocalized behavior in the ring will lead to the
$|z|\rightarrow1$ and $|z|\rightarrow0$ limits, respectively.

In addition to probing the locality of its physics, theoretical methods can also offer predictions
about an insulator's optical gap.
Although this gap was not accessible in the ground-state work of Sorella, the BLM can
directly target an excited state by minimizing the function
$\Omega=\langle\Psi|(\omega-\hat{H})|\Psi\rangle / \langle\Psi|(\omega-\hat{H})^2|\Psi\rangle$,
which, when the energy shift $\omega$ is placed inside the gap, will have the first excited state
as its global minimum \cite{Eric:2016:dir_tar}.
As this excited state approximates the state at the bottom of the infinite ring's conduction band,
this approach represents a direct, many-body, non-perturbative, and systematically improvable
route to estimating the optical gap of a solid.
In this study, we will explore a simple prototype of this approach by converging the gap for
the H$_{16}$ ring by systematically increasing the number of CSFs included in a MSJ expansion.
Although linear combinations of CSFs are not natural fits for the strongly
correlated physics of a Mott transition and will thus require a large number of CSFs be employed,
they do offer straightforward systematic improvability and anyways allows us to demonstrate
that the BLM can handle the correspondingly large number of variational parameters.

To construct our MSJ expansion, we begin by using GAMESS to optimize a (6,6) state-average
CASSCF ansatz in the cc-pVDZ basis \cite{cc-pvdz}.
We then perform a single-reference 
CISDTQ for each state, after which we truncate this expansion at
different coefficient thresholds to produce a series of increasingly large CSF
expansions.
By combining these with QMCPACK's standard spline-based, cusp-inducing $e$-$e$ and $e$-$n$
two-body Jastrow factors, we produce two sets of MSJ expansions, on each for the ground
and excited state.
Finally, choosing the value of $\omega$ that is appropriate for each state by adjusting
it to find the overall minimum of the target function $\Omega$ \cite{Eric:2016:dir_tar}, we
optimize both the CSF coefficients and Jastrow variables simultaneously using the BLM.

\begin{figure}[b]
\centering
  \includegraphics[width=0.87\linewidth,angle=270]{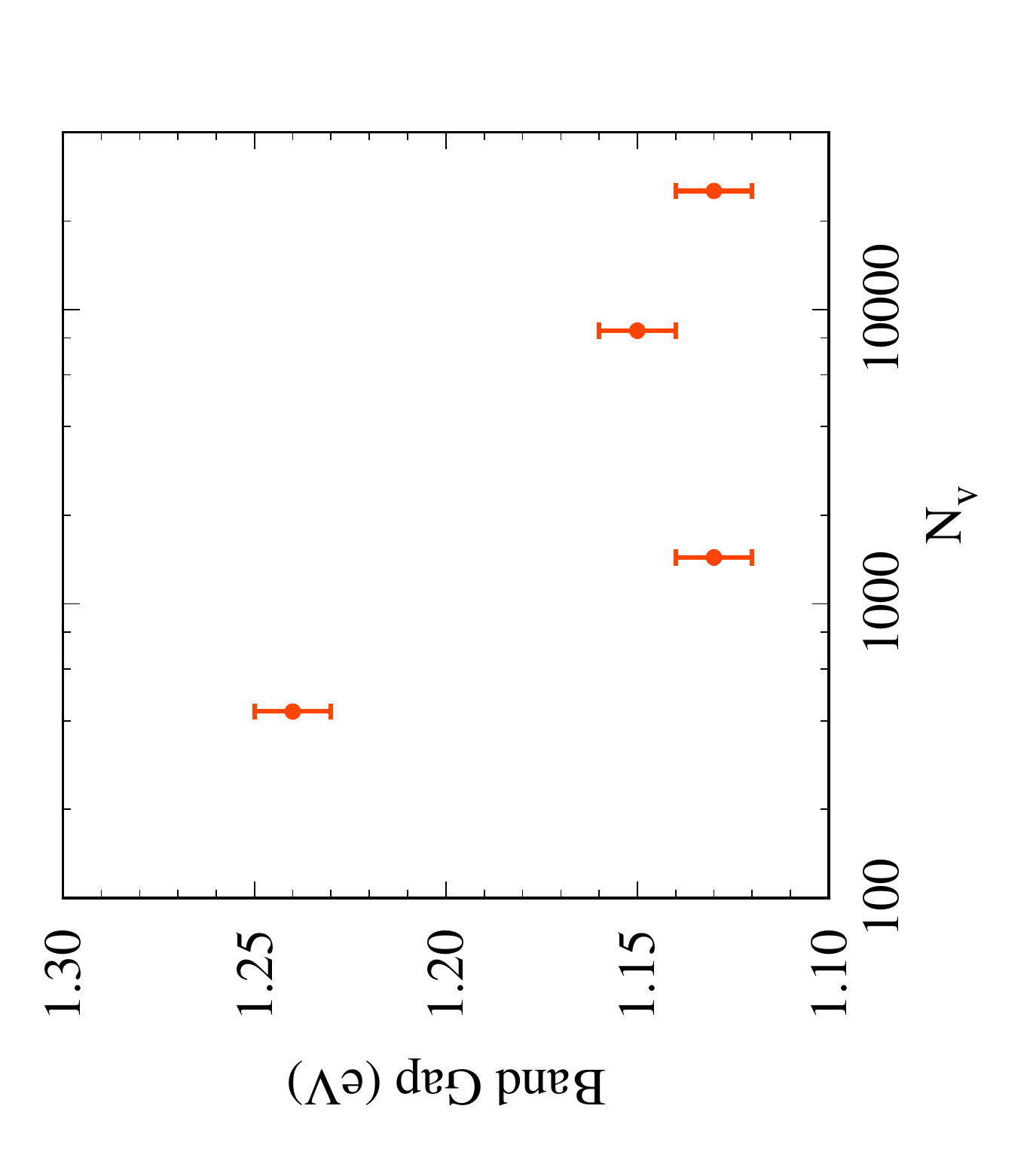}
  \caption{Convergence of the hydrogen ring's optical gap with respect to increasing variational flexibility,
           with $N_V$ the number of variational parameters in the excited state ansatz.}
\label{fig:gap_convergence}
\end{figure}

Figure \ref{fig:z_and_gap} shows the norm of the complex polarization function as well as the optical gap estimate
(defined as the difference between excited and ground state energies)
as functions of interatomic distance $a$ for a coefficient truncation threshold of 0.01.
As expected, both $|z|$ and the gap are zero for small $a$, where previous studies have found hydrogen chains
to be metallic.
As $a$ increases, we see an abrupt change in $|z|$ that suggests that by $a=3.0 a.u.$,
the ring has transitioned into an insulating state.
Being a finite system, the energy gap does not open discontinuously, and we see instead
a rapid rise in the gap until it reaches a plateau beyond $a=3.0 a.u.$,
thus agreeing with $|z|$ as to the location of the transition.

To ensure we have accurately converged the size of the gap in the insulating plateau region,
we have performed our analysis of systematically increasing CSF expansion sizes at
$a=2.95 a.u.$, where we transition from $N_B=1$ (the traditional LM) to
$(N_B=100,N_O=5,N_K=3)$ when the number of variables surpasses 5,000.
Figure \ref{fig:energy_convergence} shows the convergence behavior for the optimization of the largest
MSJ expansions for both the ground and excited states, which involved 21,401 and 25,297 variational parameters,
respectively. 
Note that, as is typical for the traditional LM, the BLM converges in a handful of iterations.
It is also important to point out that the total computational cost for evaluating all of the data points
in Figure \ref{fig:energy_convergence} amounted to 8,000 core-hours using
the 2.3 GHz Intel Xeon 12-core Haswell processors of Berkeley's Savio computing cluster.
Although this cost is not trivial, it is modest on the scale of modern parallel computation,
giving ample room for this approach to be scaled up both to larger systems and larger variational parameter sets.
Finally, in Figure \ref{fig:gap_convergence}, we show the convergence of the energy gap as the variational
flexibility of the ansatz is increased, seeing clearly that, to within our statistical uncertainty, the gap
has converged with respect to the addition of further CSFs into the wave function.
Thus, by combining the direct optimization of ansatzes for the ground and
conduction edge states with the ability to optimize the large number of parameters inherent to a systematic
expansion of ansatz flexibility, we provide an example of how the optical gap of a Mott insulator may
be converged with respect to the effects of strong, many-body correlations.

\section{Conclusions}

We have presented the blocked linear method,
a wave function optimization method for variational Monte Carlo that addresses a crucial memory bottleneck
in the highly successful traditional linear method.
By dividing ansatz variables into blocks, finding important update directions in each block, and then
combining these directions to find an overall update for the current wave function, our method
minimizes either the energy or a function suitable for targeting excited states while avoiding
both the construction of overly large matrices and any requirement that such matrices be well conditioned.
In small molecule tests that employed multiple ansatz types and involved both real space and Hilbert space
sampling, we showed that the method reproduces the results of the traditional linear method to a very good approximation.

In a demonstration of the method's ability to optimize large variable sets, we showed that the optical gap
of a Mott-insulating hydrogen ring could be systematically converged with respect to increasing flexibility
in the ansatzes for the ground and conduction band edge states.
Although there are many important concerns for real solids that did not appear in this example,
such as obtaining molecular orbitals from a density functional starting guess, addressing finite
size effects through twist averaging, and ensuring the simulation cell is sufficient to capture
excitonic effects, these issues do not present fundamental barriers and indeed have been addressed
in other contexts.
We are therefore excited to explore the new opportunities that the blocked linear method
creates in real solids and larger molecules, as well as further refinements in methodology
to bring even larger sets of variational parameters within the reach of variational Monte Carlo.

\section{Acknowledgments}
This work was supported by the U.S. Department of Energy, Office of Science, Basic Energy Sciences,
Materials Sciences and Engineering Division, as part of the Computational Materials Sciences Program.
Calculations were performed using the Berkeley Research Computing Savio cluster. 
\label{sec:acknowledgments}

\bibliographystyle{aip}

\end{document}